\documentclass[twocolumn]{aastex63}

\usepackage[T1]{fontenc}
\usepackage{aas_macros}
\usepackage{natbib}
\usepackage{times}

\newcommand{\package}[1]{\textsl{#1}}

\newcommand{\be}{\begin{equation}}
\newcommand{\ee}{\end{equation}}
\newcommand{\bea}{\begin{eqnarray}}
\newcommand{\eea}{\end{eqnarray}}

\received{2020 October 22}
\revised{2021 March 1}
\accepted{2021 March 2}

\submitjournal{ApJ Letters}
\shorttitle{Stellar Clustering and Planetary Multiplicity}
\shortauthors{Longmore, Chevance \& Kruijssen}

\usepackage{amsmath}

\begin{document}\sloppy\sloppypar\raggedbottom\frenchspacing

\title{\vspace{-8mm}The Impact of Stellar Clustering on the Observed Multiplicity and Orbital Periods of Planetary Systems}

\correspondingauthor{Steven~N.~Longmore}
\email{s.n.longmore@ljmu.ac.uk}

\author[0000-0001-6353-0170]{Steven~N.~Longmore}
\affil{Astrophysics Research Institute, Liverpool John Moores University, IC2, Liverpool Science Park, 146 Brownlow Hill, Liverpool L3 5RF, UK}

\author[0000-0002-5635-5180]{M\'{e}lanie~Chevance}
\affil{Astronomisches Rechen-Institut, Zentrum f\" ur Astronomie der Universit\"at Heidelberg, M\"onchhofstra\ss e 12-14, D-69120 Heidelberg, Germany}

\author[0000-0002-8804-0212]{J.~M.~Diederik~Kruijssen}
\affil{Astronomisches Rechen-Institut, Zentrum f\" ur Astronomie der Universit\"at Heidelberg, M\"onchhofstra\ss e 12-14, D-69120 Heidelberg, Germany}


\begin{abstract}\noindent
It has recently been shown that stellar clustering plays an important role in shaping the properties of planetary systems. We investigate how the multiplicity distributions and orbital periods of planetary systems depend on the 6D phase space density of stars surrounding planet host systems. We find that stars in high stellar phase space density environments (overdensities) have a factor 1.6--2.0 excess in the number of single planet systems compared to stars in low stellar phase space density environments (the field). The multiplicity distribution of planets around field stars is much flatter (i.e.\ there is a greater fraction of multi-planet systems) than in overdensities. This result is primarily driven by the combined facts that: (i) `hot Jupiters' (HJs) are almost exclusively found in overdensities; (ii) HJs are predominantly observed to be single-planet systems. Nevertheless, we find that the difference in multiplicity is even more pronounced when only considering planets in the \emph{Kepler} sample, which contains few HJs. This suggests that the \emph{Kepler} dichotomy -- an apparent excess of systems with a single transiting planet -- plausibly arises from environmental perturbations. In overdensities, the orbital periods of single-planet systems are smaller than orbital periods of multiple-planet systems. As this difference is more pronounced in overdensities, the mechanism responsible for this effect may be enhanced by stellar clustering. Taken together, the pronounced dependence of planetary multiplicity and orbital period distributions on stellar clustering provides a potentially powerful tool to diagnose the impact of environment on the formation and evolution of planetary systems.

\end{abstract}

\section{Introduction}
\label{sec:intro}

The comparison of planetary properties between single- and multiple-planet systems has long been used to probe the formation and evolution of planetary systems \citep{wright09, lissauer11,winn_fabrycky15, mulders18,weiss18a, weiss18b}. Motivated by the rich variety of exoplanetary systems observed with the \emph{Kepler} mission \citep{borucki10,borucki11}, the similarity of the host-star properties, planet radii, and radius valley for single-planet and multi-planet systems has been used to infer they have a common origin \citep{weiss18a,rogers20}. However, since the early days of \emph{Kepler}, it has been known that using a single population of planetary systems that matches the higher multiplicities simultaneously underpredicts the number of singly transiting systems \citep{lissauer11,hansen_murray13,ballard16}. The apparent excess of systems with a single transiting planet -- known as the \emph{Kepler} dichotomy \citep[e.g.][]{johansen12} -- places important constraints on the degree to which all planets may share a common origin. 

Different scenarios have been proposed to explain the origin of the \emph{Kepler} dichotomy \citep[see e.g.][and references therein]{he19,he20}. Some scenarios attempt to assess whether all observed planet properties can be described by a single underlying population. Such studies invoke an intrinsically high fraction of single systems \citep{fang_margot12,sandford19} or a strong anti-correlation between the mutual inclination scale and the multiplicity of each system \citep{zhu20}. Other scenarios assert that more than one planet population is required, between which the orbital properties may vary \citep[e.g.\ the mutual inclination, see][]{mulders18,he19}. Observations show an anti-correlation between planet multiplicity and their dynamical excitation \citep[e.g.][]{morton14,vaneylen19}, which points towards a scenario in which high-multiplicity systems can became dynamically unstable and lose some of their planets \citep[see e.g.][]{zinzi_turrini17}, leaving behind dynamically excited, compact systems. In this case, most observed `single' planets are, in fact, part of misaligned multiple-planet systems \citep{he20}. These scenarios suggest perturbations of planetary systems may play a role in shaping the \emph{Kepler} dichotomy.

Following the idea that the \emph{Kepler} dichotomy may result from perturbations, we assess the role that the ambient stellar clustering plays in shaping the multiplicity and orbital properties of planetary systems \citep[also see e.g.][]{cai18}. We take the sample of known exoplanets and use the ambient stellar phase space density obtained with \textit{Gaia} \citep{gaia16,gaia18} to divide the sample into low and high ambient stellar phase space densities \citep{w20nat}, which we refer to as planets residing in the `field' and in `overdensities', respectively. These subsamples are considered as reflecting environments of low and high degrees of perturbation.

This investigation mirrors those in a set of companion papers, where we investigate the impact of stellar clustering on the orbital period distribution of planets and the incidence of hot Jupiters \citep{w20nat} and on the correlation between the properties of adjacent planets \citep[i.e.\ `peas in a pod', \citealt{weiss18a}]{chevance21}, as well as its role in turning sub-Neptunes into super-Earths \citep[i.e.\ driving them across the `radius valley', \citealt{fulton17}]{kruijssen20d}. These findings demonstrate that stellar clustering has a major impact on the architectures of planetary systems, plausibly through external photoevaporation or dynamical perturbations \citep[e.g.][]{w20nat}.

In this \emph{Letter}, we investigate the extent to which the \emph{Kepler} dichotomy and the orbital properties of planets in single and multiple systems may arise from such environmental perturbations. 

We find an excess of single-planet systems in overdensities, strongly suggesting that the \emph{Kepler} dichotomy might indeed arise due to the impact of the large-scale stellar environment. The single-planet population in overdensities is characterised by shorter orbital periods than the multi-planet population, suggesting that the perturbations cause the remaining planet to migrate to tighter orbits. 

\section{Method}
\label{sec:method}

The subsequent analysis relies on the division of known exoplanetary systems into field and overdensity systems by \citet{w20nat}. The primary exoplanet sample is drawn from the `catch-all' \citet{exoarchive}, which contains a heterogeneous mix of exoplanet detections from ground-based and space-based facilities using a range of different observational techniques. To avoid issues with various observational biases related to sample selection, the \citet{w20nat} method explicitly does \emph{not} search for correlations, or look for relationships between planet properties, within the database alone. Instead, the method splits exoplanet host stars in the database in a carefully controlled way using an independent dataset (\emph{Gaia}), such that the inherent biases, selection effects, etc., in the archive are the same (within the inherent uncertainties) between the split populations. Through subsequent Monte Carlo experiments, we have verified that any residual biases when making the split are not the cause of the difference in planetary properties between the sub-samples \citep[see, e.g., Figure 3 and Extended Data Figures 3, 4, 5, 7, 8 in][]{w20nat}. 

The \citet{w20nat} sample contains 1522 planets orbiting 1137 stars, drawn from the \citet{exoarchive}. \citet{w20nat} calculated the relative, six-dimensional position-velocity phase space density of every exoplanet host star from \emph{Gaia}'s second data release \citep{gaia18} for which radial velocities are also available (1522 out of 4141 exoplanets), as well as for up to 600 neighbouring stars ($N_{\rm ss}$) within 40\,pc of each exoplanet host. For most exoplanet hosts, the resulting phase space density distributions of nearby stars is well described with a double lognormal. To quantify this, \citet{w20nat} determine the probability $P_{\rm null}$ that the phase space density distribution of each stellar host neighbourhood is drawn from a single lognormal distribution. 

As described in \citet{w20nat}, we first remove stars from this sample for which it is not possible to reliably decompose their phase space distributions into low- and high-density components. This can be either due to a low number of neighbours or because their local phase space density distribution is not bimodal (i.e.\ we remove stars for which $P_{\rm null}\ge0.05$ \& $N_{\rm ss}<400$). As exoplanet architectures correlate with the mass and age of the host star \citep{kennedy13,winn_fabrycky15}, we further only include stars with ages $1{-}4.5$\,Gyr and masses $0.7{-}2$\,M$_\odot$, to ensure that the low- and high-density sub-samples have similar distributions in these properties. These cuts leave 399 stellar systems in our fiducial reliable sample. 

For these remaining host stars, \citet{w20nat} perform a double lognormal decomposition of the local phase space density distribution, identifying a low- and a high-density component. The decomposition yields the probability $P_{\rm high}$ (respectively $P_{\rm low} = 1- P_{\rm high}$) that an individual host star lies in a phase space overdensity (respectively underdensity). We subsequently split the sample into a high ($P_{\rm high}>0.84$) and a low ($P_{\rm low}>0.84$) phase space density sample. The choice of a $1\sigma$ threshold of 0.84 in $P_{\rm high}$ and $P_{\rm low}$ represents a compromise between obtaining a large sample and minimising misclassification. The general conclusions of the paper are robust against sensible changes in this threshold.


\begin{table*}
\begin{center}
{\scriptsize
\setlength\tabcolsep{1.5mm}
\begin{tabular}{ccccccccc}
\hline
\hline
(1)                                     &   (2)          &  (3)      &  (4)   &  (5)   & (6)    & (7)    & (8)    &   (9)      \\
Sample                                 & Total number & \multicolumn{6}{c}{Number of systems ($N$) with different numbers of planets ($n_p$)}   & Ratio: sing. to mult. \\
                                             & of systems      & $N(n_p=1)$ & $N(n_p=2)$ & $N(n_p=3)$ & $N(n_p=4)$ & $N(n_p=5)$ & $N(n_p=6)$       & $N(n_p=1)/$ \\
                                                                                          &       &  &  &  &  & & & $N(n_p>1)$ \\

\hline

Full reliable sample              &    399   & 311 & 68 &  9 &  7 &  3 &  1 & 3.53$\pm$0.43 \\
(`All')                                     &             &        &     &     &     &     &     &         \\
\hline
Low density                          &      48    & 35   & 11 &  2 &  0   & 0  &  0 & 2.69$\pm$0.87 \\
(`Field')                                  &             &        &     &     &     &     &     &         \\
\hline
High density                         &    253   & 205 & 36 &  6 &  4  & 2 &  0  & 4.27$\pm$0.68 \\
(`Overdensities')                  &             &        &     &     &     &     &     &         \\
\hline
Low density, no hot Jupiters              &    40   & 27 & 11 &  2 &  0 &  0 &  0 & 2.08$\pm$0.70 \\
(`Field, no HJ')                                     &             &        &     &     &     &     &     &         \\
\hline
High density, no hot Jupiters      &      157    & 112   & 33 &  6 &  4   & 2  &  0 & 2.48$\pm$0.44 \\
(`Overdensities, no HJ')                                  &             &        &     &     &     &     &     &         \\
\hline
Low density, Kepler \& K2                &    8   & 5 & 2 &  1 &  0 &  0 &  0 & 1.67$\pm$1.22 \\
(`Field, Kepler/K2')                                     &             &        &     &     &     &     &     &         \\
\hline
High density, Kepler \& K2              &      79    & 52   & 19 &  5 &  3   & 0  &  0 & 1.93$\pm$0.46 \\
(`Overdens, Kepler/K2')                                  &             &        &     &     &     &     &     &         \\
\hline
Low density, Kepler-only                &    8   & 5 & 2 &  1 &  0 &  0 &  0 & 1.67$\pm$1.22 \\
(`Field, Kepler')                                     &             &        &     &     &     &     &     &         \\
\hline
High density, Kepler-only              &      72    & 46   & 18 &  5 &  3   & 0  &  0 & 1.77$\pm$0.43 \\
(`Overdens, Kepler')                                  &             &        &     &     &     &     &     &         \\
\hline
Low density, WASP$+$HAT         &    4   & 4 & 0 &  0 &  0 &  0 &  0 & - \\
(`Field, WASP$+$HAT')                                     &             &        &     &     &     &     &     &         \\
\hline
High density, WASP$+$HAT              &      58    & 56   & 2 &  0 &  0   & 0  &  0 & 28.00$\pm$20.15 \\
(`Overdens, WASP$+$HAT')                                  &             &        &     &     &     &     &     &         \\
\hline\hline
Low density, Kepler (CKS)         &    5   & 2 & 2 &  0 &  1 &  0 &  0 & 0.67$\pm$0.61 \\
(`Field, Kepler/CKS')                                     &             &        &     &     &     &     &     &         \\
\hline
High density, Kepler (CKS)              &      60    & 38   & 13 &  5 &  3   & 0  &  1 & 1.73$\pm$0.46 \\
(`Overdens, Kepler/CKS')                                  &             &        &     &     &     &     &     &         \\
\hline
Low density, Kepler (CKS),         &    6   & 2 & 3 &  0 &  1 &  0 &  0 & 0.50$\pm$0.43 \\
M$_*>0.5$\,M$_\odot$                                     &             &        &     &     &     &     &     &         \\
(`Field, Kep/CKS,0.5\,M$_\odot$')                                     &             &        &     &     &     &     &     &         \\
\hline
High density, Kepler (CKS),               &      69    & 44   & 15 &  5 &  4   & 0  &  1 & 1.76$\pm$0.44 \\
M$_*>0.5$\,M$_\odot$                                  &             &        &     &     &     &     &     &         \\
(`Overdens, Kep/CKS,0.5\,M$_\odot$')                                  &             &        &     &     &     &     &     &         \\
\hline
Low density, Kepler (CKS),         &    6   & 2 & 3 &  0 &  1 &  0 &  0 & 0.50$\pm$0.43 \\
0.5\,M$_\odot \, <$M$_*<1.0$\,M$_\odot$                                     &             &        &     &     &     &     &     &         \\
(`Field, Kep/CKS,$0.5-1.0$\,M$_\odot$')                                     &             &        &     &     &     &     &     &         \\\hline
High density, Kepler (CKS),               & 27  & 18   & 7 & 0  & 1    & 0  & 1  & 2.00$\pm$0.82  \\
0.5\,M$_\odot \, <$M$_*<1.0$\,M$_\odot$                                  &             &        &     &     &     &     &     &         \\
(`Overdens, Kep/CKS,$0.5-1.0$\,M$_\odot$')                                  &             &        &     &     &     &     &     &         \\

\hline

\hline
\end{tabular}
}
\caption{The number of stellar systems and the planet multiplicity distributions for different data samples. The second column gives the total number of planetary systems in each of the samples. Columns 3 to 8 give the number of systems in each of the samples with the specified number of detected planets. The final column gives the ratio (and associated Poisson uncertainty) of the number of single and multiple planetary systems. All samples above the double horizontal line use the stellar parameters from the NASA Exoplanet Archive \citep{exoarchive}. The Kepler-only samples below the double horizontal line use stellar parameters from the California-Kepler Survey \citep[CKS, taken from][]{fulton_petigura18}. Rows with M$_*>0.5$\,M$_\odot$ use a lower mass cutoff of $0.5$\,M$_\odot$ for the host stellar mass, rather than $0.7$\,M$_\odot$ as for all other samples. The final row uses an upper mass cutoff of 1.0\,M$_\odot$, rather than 2.0\,M$_\odot$ for all other samples. The WASP$+$HAT rows are included to show that ground-based data currently do not have large enough samples of planets in low-density environments to enable a meaningful comparison of multiplicities.}
\label{tab:multiplicity}
\end{center}
\end{table*}


\begin{table*}
\begin{center}
{\footnotesize
\begin{tabular}{cccccccc}
\hline
\hline
(1)                                     &   (2)          &  (3)      &  (4)   &  (5)   & (6)    & (7)    & (8)        \\
Sample                                 & Stellar  & Stellar      & Stellar  & Distance   & Number & Num. systems            & Num. systems   \\
                                             &    mass  & metallicity &   age    &       from sun            & of     & with $\ge 1$ transit       & with $\ge 1$ radial            \\
                                             & [M$_\odot$]  &  [dex] &  [Gyr]  &  [pc]  & systems &  detections       &  velocity detections          \\

\hline
                                             &             &        &     &     &     &     &              \\
Low density                          & $1.09_{-0.24}^{+0.41}$ & $0.08_{-0.28}^{+0.18}$ & $3.2_{-1.3}^{+0.6}$ & $65.0_{-33.0}^{+197.0}$ & $48$ & $17$ & $40$ \\
(`Field')                                 &             &        &     &     &     &     &              \\
\hline
                                             &             &        &     &     &     &     &              \\
High density                         &   $1.2_{-0.25}^{+0.21}$ & $0.09_{-0.13}^{+0.15}$ & $2.7_{-1.0}^{+1.3}$ & $218.0_{-167.0}^{+214.0}$ & $253$ & $160$ & $182$ \\
(`Overdensities')                  &             &        &     &     &     &     &              \\

\hline
                                            &             &        &     &     &     &     &              \\
Low density, Kepler (CKS)                  & $0.78_{-0.05}^{+0.35}$  & $0.02_{-0.33}^{+0.15}$ & $3.8_{-0.6}^{+0.5}$ & $172.0_{-19.0}^{+60.0}$ & $5$ & $5$ & $1$  \\
(`Field, Kepler/CKS')                                 &             &        &     &     &     &     &              \\
\hline
                                             &             &        &     &     &     &     &              \\
High density, Kepler (CKS)              &  $1.08_{-0.14}^{+0.18}$  & $0.07_{-0.1}^{+0.13}$ & $2.9_{-1.0}^{+1.1}$ & $389.0_{-144.0}^{+137.0}$ & $60$ & $60$ & $14$ \\
(`Overdens, Kepler/CKS')                  &             &        &     &     &     &     &              \\

\hline
                                             &             &        &     &     &     &     &              \\
Low density, Kepler (CKS),                  & $0.75_{-0.06}^{+0.33}$ & $-0.12_{-0.19}^{+0.22}$ & $3.8_{-0.6}^{+0.5}$ & $172.0_{-19.0}^{+60.0}$ & $6$ & $6$ & $1$ \\
M$_*>0.5$\,M$_\odot$                       &  &  &  &  &  &  &  \\
(`Field, Kepler/CKS,0.5\,M$_\odot$')       &             &        &     &     &     &     &              \\
\hline
                                             &             &        &     &     &     &     &              \\
High density, Kepler (CKS),              & $1.07_{-0.13}^{+0.18}$   & $0.07_{-0.1}^{+0.12}$ & $2.7_{-1.4}^{+1.3}$ & $354.0_{-121.0}^{+172.0}$ & $69$ & $69$ & $15$ \\
M$_*>0.5$\,M$_\odot$                       &  &  &  &  &  &  &  \\
(`Overdens, Kepler/CKS,0.5\,M$_\odot$')   &             &        &     &     &     &     &              \\

\hline
                                             &             &        &     &     &     &     &              \\
High density, Kepler (CKS),              & $0.94_{-0.12}^{+0.04}$ & $0.02_{-0.06}^{+0.07}$ & $1.7_{-0.7}^{+1.5}$ & $285.0_{-96.0}^{+73.0}$ & $27$ & $27$ & $6$ \\
$0.5\,$M$_\odot<\,$M$_* \, <1.0$\,M$_\odot$                       &  &  &  &  &  &  &  \\
(`Overdens, Kepler/CKS, $0.5-1.0$\,M$_\odot$')   &             &        &     &     &     &     &              \\

\hline \hline

\end{tabular}
}
\caption{Investigation of potential observational biases in the properties of different field and overdensity samples (see Table~\ref{tab:multiplicity} for sample details). Columns 2 to 5 show the mean~$\pm1\sigma$ distribution of host stellar mass, host stellar metallicity, host stellar age and distance from the Sun. Columns 6 to 8 show the total number of stellar systems and the number of systems in which at least one planet has been detected through transit and radial velocity measurements, respectively.}
\label{tab:bias_check}
\end{center}
\end{table*}

Table~\ref{tab:multiplicity} shows the number of stars and the planet multiplicity distributions for the full reliable sample and different low- and high-density sub-samples, which we refer to as the `field' and `overdensities', respectively.  Table~\ref{tab:bias_check} shows the mean distribution of host stellar mass, host stellar metallicity, host stellar age, distance from the Sun and the number of systems in which at least one planet has been detected through transit and radial velocity (RV) measurements, for the sub-set of the samples in Table~\ref{tab:multiplicity} with statistically significant differences in the single-to-multiple planet ratios between the field and overdensities.

\section{Results}
\label{sec:results}

We break up the analysis into three parts. Firstly, in $\S$\ref{sub:mult_full} and 
$\S$\ref{sub:rob_res_full} we focus on planetary multiplicity, as well as on their orbital periods and resonances, respectively, using the full \citet{w20nat} sample (i.e. the `Field' and `Overdensities' samples in the top rows of Tables~\ref{tab:multiplicity} and \ref{tab:bias_check}). The \citet{w20nat} catalogue represents the largest possible sample of field and overdensities, at the expense of the highest degree of sample heterogeneity in terms of e.g.\ selection and detection method. In $\S$\ref{sub:mult_subsamps}, we focus on sub-samples of the \citet{w20nat} catalogue that are drawn from individual exoplanet surveys to reduce the data heterogeneity, at the expense of sample size.

\subsection{Multiplicity: full \citet{w20nat} sample}
\label{sub:mult_full}

We start by investigating the multiplicity of the `Field' and `Overdensities' samples from the full \citet{w20nat} catalogue. \cite{w20nat} ruled out stellar mass, metallicity and distance as systematic sources of bias between the field and overdensity samples. It is therefore unlikely that observational biases in these parameters would manifest themselves as differences in planet multiplicity (or orbital period (ratios), $\S$\ref{sub:rob_res_full}) between the field and overdensity samples.

\begin{figure*}
\includegraphics[width=\hsize]{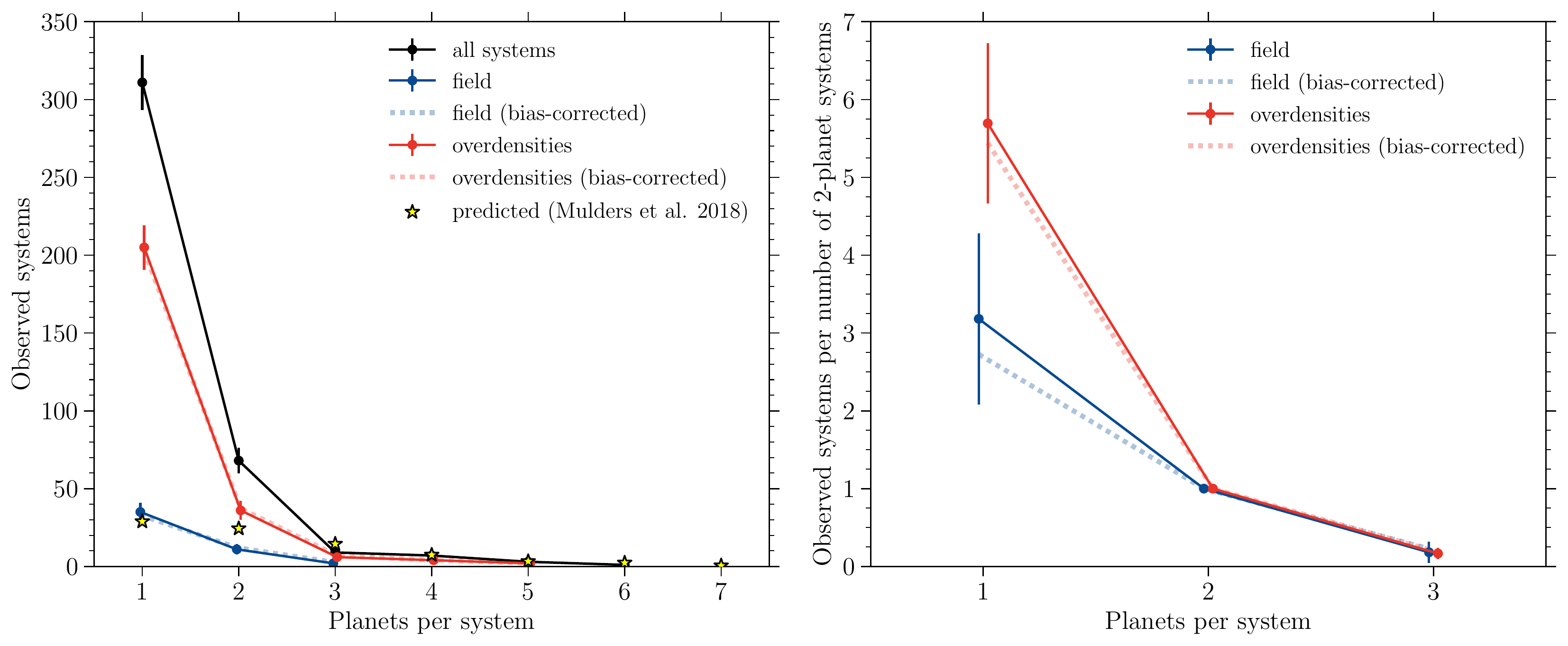}%
\vspace{-2mm}\caption{
\label{fig:planets_per_system}
Planetary system multiplicity distributions from the full \citet{w20nat} sample. The left panel shows multiplicity distributions for the full reliable sample (black, `all systems'), as well as the low-density (blue, `field'), and high-density (red, `overdensities') sub-samples. The shaded blue and red dotted lines show the values corrected for the method detection bias (see the text). Error bars represent the Poissonian ($\sqrt N$) uncertainties on the data points. Stars show the co-planar multiplicity distribution predicted by \citet[their fig.~11]{mulders18}, scaled down by a factor of 30 to enable an easier comparison to the samples in this work.  The right panel shows the same planet multiplicity distributions, but this time each sample has been normalised to the number of systems with two planets. Both panels show that the planet multiplicity distribution of systems in the `field' and `overdensities' differ considerably. While overdensities are greatly dominated by single-planet systems, the planet multiplicity distribution of stars in the field sample is significantly flatter and more closely matches the model predictions.}

\end{figure*}

Table~\ref{tab:multiplicity} shows there are clear differences in the distribution of planet multiplicity between the field and overdensities. In particular, overdensities have a greater fraction of systems with only one detected planet. The final column of Table~\ref{tab:multiplicity} shows overdensities have a factor 1.6 excess of single planet systems compared to the field. Figure~\ref{fig:planets_per_system} visualises the planet multiplicity distributions of both sub-samples. The planet multiplicity distribution of stars in the field sample is significantly flatter, i.e.\ a greater fraction of systems in the field have more than one planet. 

Table~\ref{tab:bias_check} also shows that there are differences in the fraction of planets in the field and overdensities that have been detected through transit and RV observations. The most notable difference is that only 35\% of the field star systems have planets detected through transits compared to 63\% for overdensities. Given that transit and RV observations are sensitive to detecting planets in different mass and orbital period regimes, it is possible that this imbalance in detection method may imprint a bias in the planetary multiplicity, orbital period and orbital period ratio between the field and overdensity samples.  


To estimate the potential bias due to imbalances in observational detection methods, we split the overdensity\footnote{A similar experiment cannot be performed for the field sample due to the low number of field systems.} sample into systems where all planets are exclusively detected through transits (transit-only sample) and those where all planets are exclusively detected through radial velocity measurements (RV-only sample). These sub-samples contain 67 and 89 systems, respectively. We repeat the entire analysis in $\S$\ref{sub:mult_full} and $\S$\ref{sub:rob_res_full} using the transit-only and RV-only samples to determine how robust the identified differences between the full \citet{w20nat} field and overdensity samples are against detection method bias. We then focus our discussion on results that are found to be robust, defined such that any bias would act in the opposite direction of the identified trends, and that correcting for this bias would either strengthen the results or leave them unchanged, rather than weakening them. We explicitly refer to the results of the detection method bias test after each analysis step below.


Repeating the multiplicity analysis on the RV-only and transit-only samples shows that the RV-only samples have a steeper drop in multiplicity, i.e.\ they are more likely to detect single-planet systems. As the field sample planet population is dominated by RV measurements and recovers a smaller fraction of single-planet systems than overdensities, we infer that the difference in multiplicity distribution seen between the field and overdensity systems cannot be explained by a detection method bias, and that accounting for this bias would only strengthen the above results.

\begin{figure*}
\includegraphics[width=\hsize]{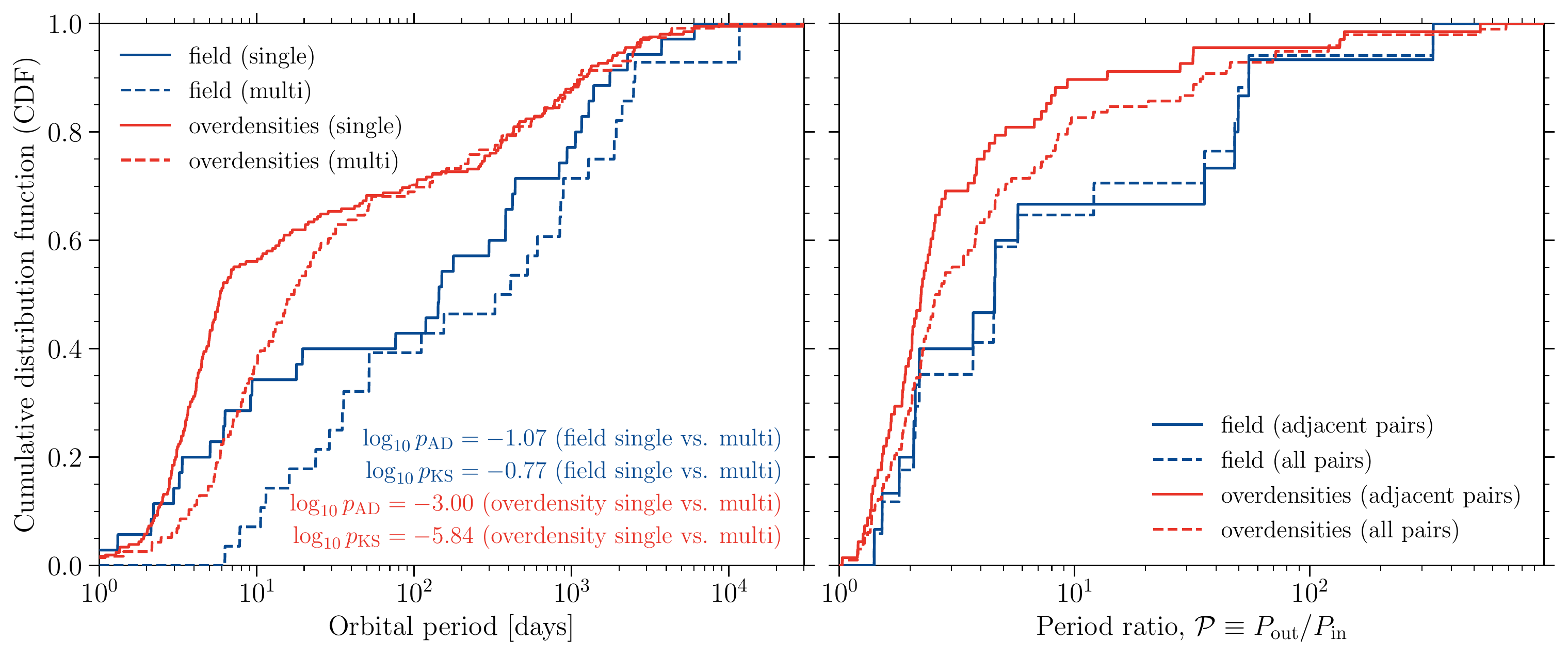}%
\vspace{-2mm}\caption{
\label{fig:planets_cdf_period_ratio}
Orbital period demographics in field and overdensity systems for the full \citet{w20nat} sample. The left panel shows the cumulative distribution functions of orbital periods for planets around stars in overdensities (red) and the field (blue). The solid lines show the distribution for stellar systems with only a single planet (`single'), while the dashed lines show the distribution for stellar systems with more than one planet (`multi'). The $p_{\rm KS}$ and $p_{\rm AD}$ values in the bottom right corner show the results of two-sample Kolmogorov-Smirnov and Anderson-Darling tests, respectively, assessing the null hypothesis that the two samples listed in parentheses after the $p_{\rm KS}$ and $p_{\rm AD}$ values are drawn from the same population. This is ruled out at high confidence for the overdensity sample, for which the orbital periods of single-planet systems are significantly smaller than those of multiple-planet systems. For the field sample, the period distributions of single and multi-planet systems are statistically indistinguishable. The right panel shows the cumulative distribution function of pairs of planets within the same system with period ratio $\mathcal{P} = P_{\rm out}/P_{\rm in} $, where $P_{\rm out}$ and $P_{\rm in}$ are the periods of the outer and inner planet, respectively. Dashed lines show all planet pairs and solid lines show only adjacent planet pairs. Blue and red show the field and overdensity samples, respectively. Planetary systems in overdensities may be somewhat more closely packed than in the field, but not at high significance. }
\end{figure*}

To illustrate this point, we formulate an approximate correction of the field and overdensity samples for the detection method bias. We measure the number of systems per multiplicity value in the RV-only and transit-only samples, and normalise them by the total number of systems in the RV-only sample ($N_{\rm ov,rv} = 89$) and in the transit-only sample ($N_{\rm ov,tr} = 67$). For a given multiplicity value, $i$, the normalised fraction of systems is then $f_{i,\rm rv}$ and $f_{i,\rm tr}$ for the RV-only and transit-only samples, respectively. Relative to a sample containing an equal number of transit and radial velocity detections, the bias of each individual detection method in terms of the fraction of systems per multiplicity value is defined as, 
\be
\label{eq:bias}
\delta_{i,\rm rv} = \frac{1}{2}\left(f_{i,\rm rv}-f_{i,\rm tr}\right) ~~~{\rm and}~~~
\delta_{i,\rm tr} = \frac{1}{2}\left(f_{i,\rm tr}-f_{i,\rm rv}\right) .
\ee
We then weigh these biases by the numbers of systems that are detected only by radial velocities ($N_{\rm rv}$) or transit ($N_{\rm tr}$) in each of the field and overdensity samples to correct the measurements as
\be
\label{eq:corr}
f_{i}\rightarrow f_{i}-\left(\frac{N_{\rm rv}}{N_{\rm rv}+N_{\rm tr}}\delta_{i,\rm rv}+\frac{N_{\rm tr}}{N_{\rm rv}+N_{\rm tr}}\delta_{i,\rm tr}\right).
\ee
The number of systems per multiplicity value $i$ is then this corrected fraction $f_i$ multiplied by the total number of systems in the sample (field or overdensity). The results are shown as shaded dotted lines in \autoref{fig:planets_per_system} and clearly show that correcting for detection method biases would strengthen our finding that overdensities show an excess of single-planet systems compared to the field, increasing the relative excess of single-planet systems in overdensities relative to the field from 1.6 to 2.0. This suggests that future work repeating this analysis on larger and more homogeneous samples has the potential to find an even stronger multiplicity distribution difference between systems in the field and in overdensities.


\subsection{Orbital period and resonances: full sample}
\label{sub:rob_res_full}

We now turn to a discussion of the orbital period distributions as a function of planetary multiplicity for the full \citet{w20nat} sample. The left-hand panel of Figure~\ref{fig:planets_cdf_period_ratio} shows the cumulative distribution functions of the orbital period for the field and overdensities, split into systems with only one planet (`single') and more than one planet (`multi'). We conduct two-sample Anderson-Darling (AD) and Kolmogorov-Smirnov (KS) tests assessing whether the orbital period distributions of both single- and multi-planet systems in the field and overdensities are drawn from the same population. As reported in \citet{w20nat}, the orbital periods of planets in the field are significantly larger than in overdensities. The new analysis here shows that this holds for both single and multiple planet systems. However, repeating this analysis with RV-only and transit-only samples shows that there is a strong bias in the RV-only sample towards detecting planets with longer orbital periods. As planets detected around field stars are dominated by RV observations, the observed increase in orbital period for planets in the field compared to overdensities may be affected by detection method bias. We therefore refrain from a quantitative comparison of the orbital period distributions between the field and overdensity samples.

Instead, we focus on comparing the orbital period distributions between the single- and multiple-planet samples within the field and overdensities independently. This ensures the analysis is not affected by detection method bias. The resulting $p_{\rm AD}$ and $p_{\rm KS}$-values, shown in Figure~\ref{fig:planets_cdf_period_ratio}, strongly rule out the null hypothesis that the orbital periods of single and multiple planet systems in overdensities are drawn from the same parent distribution. In overdensities, the observed orbital periods of single-planet systems are significantly smaller than orbital periods of multiple-planet systems. While the orbital period distributions of single-planet systems in the field also appear smaller than for multiple-planet systems, the $p_{\rm AD}$ and $p_{\rm KS}$-values cannot rule out that they are drawn from the same population. 

For all systems with more than one planet, we then derive the ratio of orbital periods, $\mathcal{P}$, of the outer ($P_{\rm out}$) to inner ($P_{\rm in}$) planet, $\mathcal{P} \equiv P_{\rm out}/P_{\rm in}$,  for all planet pairs within that system. Figure~\ref{fig:planets_cdf_period_ratio} shows the cumulative distribution function of $\mathcal{P}$ for the field and overdensity samples, for all planet pairs and adjacent planet pairs. 

The period ratio for planet pairs in overdensities appears systematically smaller than for the field, with a slightly more pronounced offset for neighbouring planets than for all planet pairs. We conduct a two-sample KS test against the null hypothesis that the field and overdensity samples are drawn from the same $\mathcal{P}$ distribution. Despite the apparent offset between the field and overdensity period ratios, the resulting $p_{\rm KS}$ ($p_{\rm AD}$) values of 0.14 (0.06) for adjacent pairs and 0.31 (0.25) for all pairs mean the offset is not statistically significant enough to rule out the null hypothesis. In addition, repeating the analysis with the RV-only and transit-only samples suggests a small potential bias of RV-only samples towards longer period ratios. \citet{lissauer11} report a similar offset in their comparison of \emph{Kepler} and RV samples. As the field planet population is dominated by RV measurements, the apparent increase in $\mathcal{P}$ compared to overdensities may therefore be affected by detection method bias.


\begin{figure*}
\includegraphics[width=\hsize]{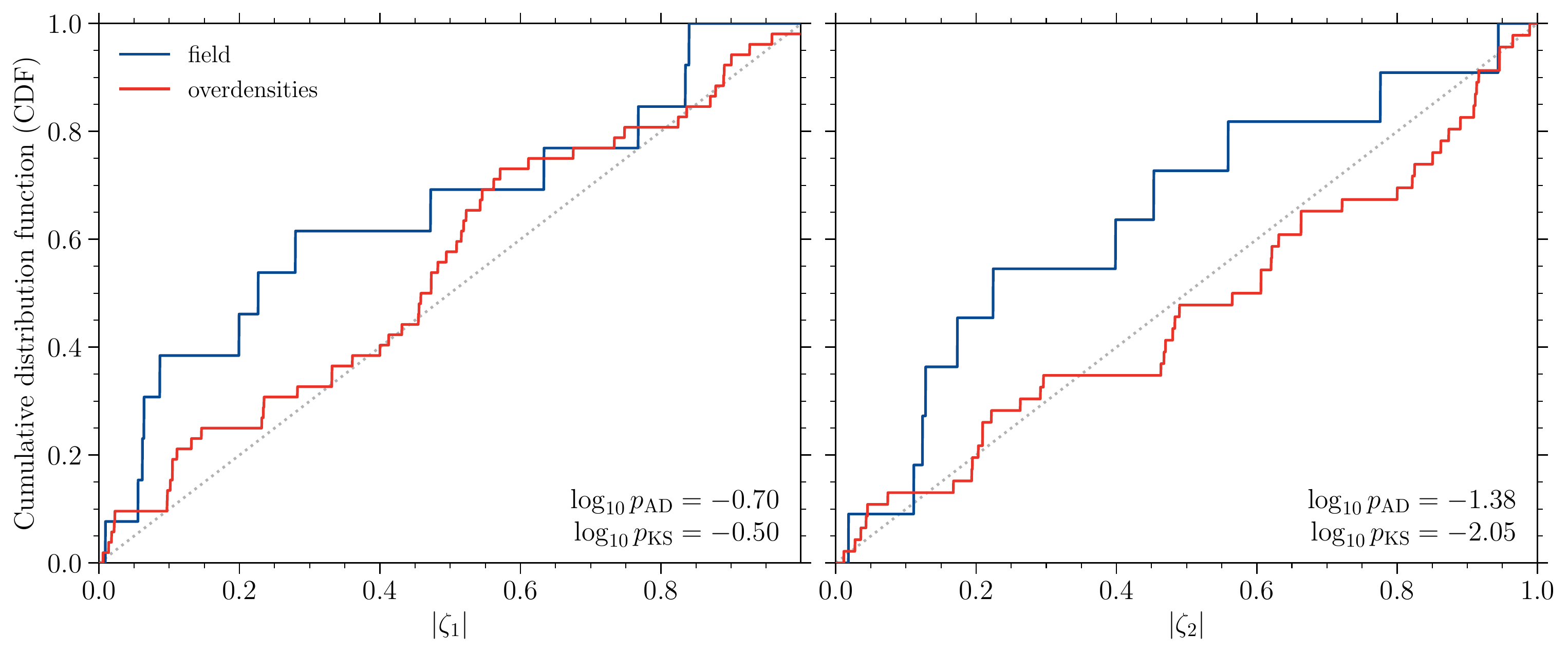}%
\vspace{-2mm}\caption{
\label{fig:zetas_adj}
Cumulative distribution functions comparing the orbital period ratios of adjacent planet pairs to mean motion resonances (MMRs) for the field (blue) and overdensity (red) sub-samples from the full \citet{w20nat} catalogue. The variables $\zeta_1$ (left panel) and $\zeta_2$ (right panel) quantify the difference between an observed period ratio and first-order and second-order MMRs, respectively, with $\zeta=0$ indicating that a pair coincides with an MMR. The dotted grey line shows the 1:1 relation between the CDF and $\zeta$ values for reference. The bottom-right corner of each panel shows the $p_{\rm KS}$ and $p_{\rm AD}$ values for two-sample Kolmogorov-Smirnov and Anderson-Darling tests, respectively, against the null hypothesis that the field and overdensity samples are drawn from the same $\zeta$ distributions. There is no statistically significant difference between the $\zeta_1$ distributions of the field and overdensity samples. However, the $\zeta_2$ distributions of the field and overdensities are not drawn from the same population. The orbital period ratios of adjacent planets in the field sample statistically lie closer to second-order MMRs than in the overdensity sample.
}
\end{figure*}

Finally, we investigate whether there are any differences in the distribution of orbital period ratios of planetary system pairs in relation to mean motion resonances (MMRs; orbital period ratios that are nearly equal to ratios of small integers) between the field and overdensity samples. Following \citet{lissauer11}, we use the variable $\zeta$ to measure the difference between an observed period ratio and nearby MMRs. In order to treat all neighbourhoods equally, $\zeta$ ranges between $-1$ and $1$ in each neighbourhood. 

For first-order MMRs (for which the orbital period ratios are $j$:$j-1$, i.e.\ 2:1, 3:2, 4:3, $\ldots$), $\zeta_1$ is given by
\be
\label{eq:z1}
\zeta_1 = 3 \left[ \frac{1} {\mathcal{P} - 1} - {\rm Round} \left (\frac{1} {\mathcal{P} - 1}\right) \right],
\ee
where `Round' returns the nearest integer. For second order MMRs (for which the orbital period ratios are $j$:$j-2$, i.e., 3:1, 4:2, 5:3, $\ldots$), $\zeta_2$ is given by the analogous expression
\be
\label{eq:z2}
\zeta_2 = 3 \left[ \frac{2} {\mathcal{P} - 1} - {\rm Round} \left (\frac{2} {\mathcal{P} - 1}\right) \right].
\ee

We conduct two-sample KS and AD tests against the null hypothesis that the field and overdensity samples are drawn from the same  $\zeta_1$ and $\zeta_2$ distributions. When including all planet pairs, there are no statistically significant differences in either the $\zeta_1$ or $\zeta_2$ distributions between the field and overdensity samples. However, when only considering adjacent planet pairs, we do find a statistically significant difference.

Figure~\ref{fig:zetas_adj} shows cumulative distribution functions of the relative proximity of the orbital period ratios of adjacent planet pairs to MMRs, for the field and overdensity sub-samples. The $p_{\rm KS}$ ($p_{\rm AD}$) value of 0.32 (0.20) shows that there is no statistically significant difference between the $\zeta_1$ distributions of the field and overdensity samples. Given our null hypothesis threshold, $p_{\rm ref}=0.05$, the $p_{\rm KS}$ ($p_{\rm AD}$) value of $8.9\times10^{-3}$ (0.04) suggests that the $\zeta_2$ distributions of the field and overdensities are not likely to be drawn from the same population. However, we need to correct for the fact that we are searching for multiple correlations within the data, which increases the chances of a false positive result.

We use the Holm-Bonferroni (H-B) method (\citealt{holm79}; see Appendix~B of \citealt{kruijssen19} for a recent astrophysical application) to check whether the difference in $\zeta$ distributions are robust against random fluctuations when searching for multiple different correlations within a dataset. We split the $p$-values shown in Figures~\ref{fig:planets_cdf_period_ratio} and \ref{fig:zetas_adj} by test statistic, order the four $p$-values in increasing value, and label them $i=1,..,4$. We then test each $p$-value in increasing $i$ order against the H-B criterion
\begin{equation}
\label{eq:hb}
p_i \leq \frac{p_{\rm ref}}{N_{\rm samp} +1 - i} \,\,\,\,\, ,
\end{equation}
where $N_{\rm samp}=4$ is the number of samples and $p_{\rm ref}=0.05$. In cases where the above condition is satisfied, the null hypothesis (that both distributions are drawn from the same underlying sample) can be rejected.

For the KS tests, the first two $p$-values ($p_1 = 1.4\times10^{-6} \leq 0.0167$, $p_2 = 8.9\times10^{-3} \leq 0.0125$) pass the H-B criterion, so both the previously reported orbital period distributions and the $\zeta_2$ distributions are statistically different. Repeating this for the AD tests ($p_1 = 0.001 \leq 0.0125$, $p_2= 0.04 \nleq 0.0167$), we find that the orbital period distributions pass the H-B criterion but the $\zeta_2$ distributions do not.

Given the $\zeta_2$ distributions are statistically different with one test statistic (KS) but not the other (AD), we opt to report the result as `marginal'. Future work with improved data is required to determine whether the orbital period ratios of adjacent planets in the field sample statistically lie closer to second-order MMRs than in the overdensity sample.  We note that repeating the analysis on the RV-only and transit-only samples shows that the detection method bias works in the opposite direction to the observed trend. In other words, correcting for this bias would likely strengthen the result that the $\zeta_2$ distributions are not drawn from the same underlying sample population.

In summary, the proximity of adjacent planet pairs to first-order MMRs does not differ significantly between field and overdensity systems. A larger sample is needed to determine whether adjacent planet pairs in overdensities are found near second-order MMRs significantly less often than those in the field.

\subsection{Multiplicity: individual exoplanet surveys}
\label{sub:mult_subsamps}

Finally, we return to investigating the difference in exoplanetary multiplicity between the systems in the field and overdensities, but now using sub-sets of the \citet{w20nat} data. The goal of this exercise is to address the heterogeneity of the sample.

Rows 4 and 5 in Table~\ref{tab:multiplicity} show the exoplanet multiplicity distributions of the field and overdensities when removing systems with hot Jupiters (HJs; here defined by masses $>50$\,M$_\earth$ and semi-major axes $<0.2$\,au) from the \citet{w20nat} sample. The single-to-multiple ratio of the field and overdensities become statistically indistinguishable. This shows that the larger single-to-multiple planet ratio in overdensities compared to the field in the full sample is a direct consequence of HJs being (i)  almost exclusively found in overdensities \citep{w20nat}, and (ii) unlikely to have close companions \citep{steffan12}.

To see if this can fully explain the difference in multipicity between the field and overdensities, we then concentrate on data from the \citet{w20nat} sample that are drawn from a single observational survey. This approach has several advantages. Firstly, the detection method for all sources is identical, so detection biases (such as RV versus transit) are not an issue. Secondly, the sample selection for the parent populations is identical. However, these advantages are offset by the fact that the number of planets in the field and overdensity samples are  reduced.

Table~\ref{tab:multiplicity} shows that no matter which individual survey is chosen, the single-to-multiple ratio is higher in overdensities than the field. However, the low number (or lack) of low density systems means that it is not possible to tell if this difference is statistically significant.

To investigate whether the lack of statistical significance might be caused by poorly constrained stellar parameters, we repeated the analysis using only the \emph{Kepler} planet data, and replacing the stellar parameters in the NASA Exoplanet Archive with those from the California-Kepler Survey \citep[][denoted by `CKS' in Table~\ref{tab:multiplicity} and \ref{tab:bias_check}]{fulton_petigura18}. We find that the single-to-multiple ratio contrast increases substantially, with overdensities having a factor 2.6 larger single-to-multiple ratio than the field. This increases to a factor of 3.5 when the lower mass cutoff of the stars is reduced from 0.7\,M$_\odot$ (as in \citealt{w20nat}) to 0.5\,M$_\odot$, which ensures that all Kepler stars are included.


\begin{figure}
\includegraphics[width=\hsize]{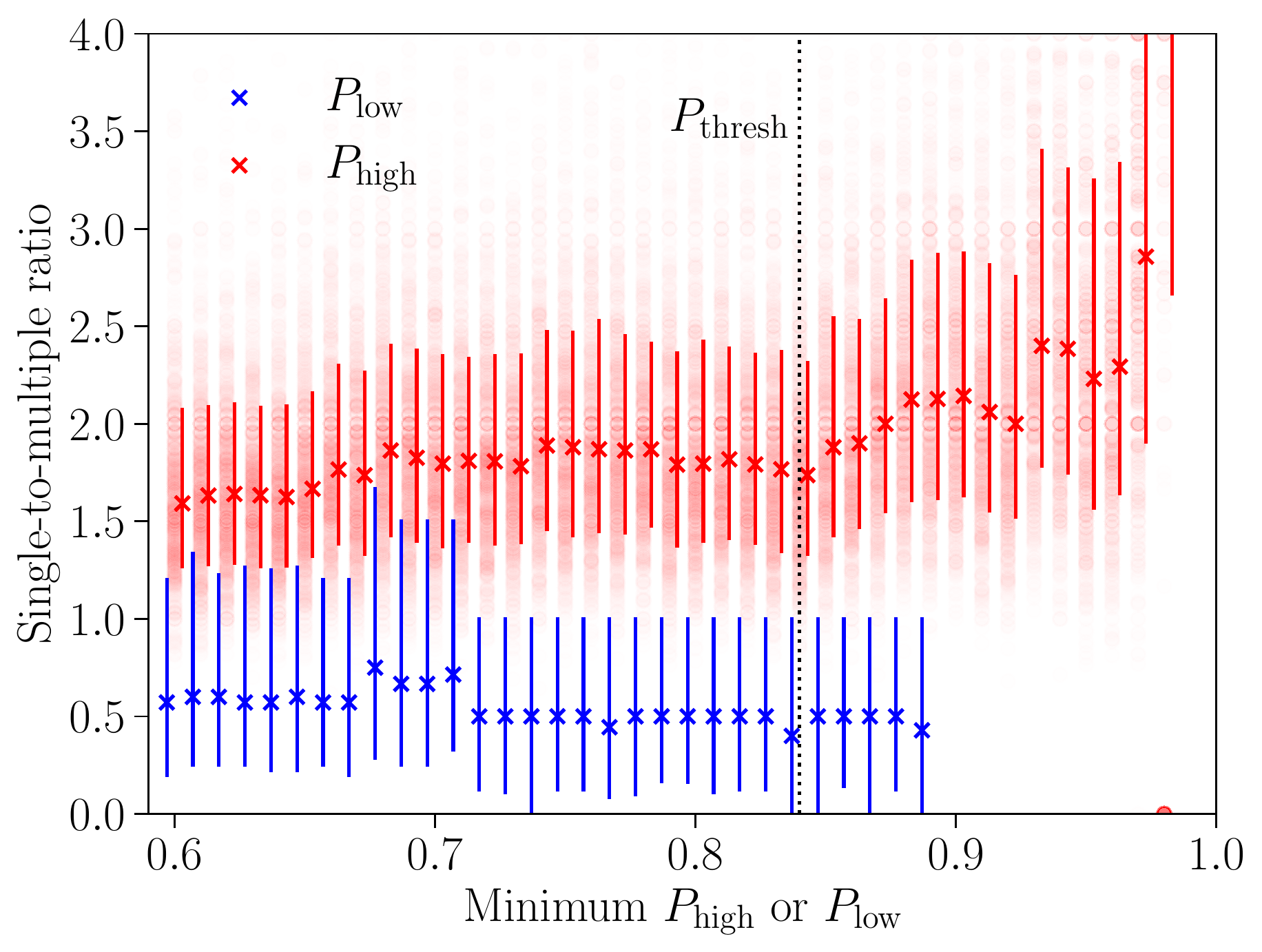}%
\vspace{-2mm}
\caption{\label{fig:Kepler_multiplicity}
Ratio of single to multiple planet systems detected with \emph{Kepler} as a function of the probability that the host star is in a high density ($P_{\rm high}$, red) or low density ($P_{\rm low}$, blue) environment. The crosses show the median single-to-multiple ratio at each $P_{\rm high}$ and $P_{\rm low}$ from 1000 Monte Carlo samples of the data drawn using Poisson statistics from the observed multiplicity. The error bars show the 16$^{\rm th}$ to 84$^{\rm th}$ percentile range of the Monte Carlo samples. The red, opacity-weighted circles show single Monte Carlo realisations for the high density sample. The vertical dotted line shows the fiducial threshold value of $P=0.84$ used to assign systems to overdensities or the field. 
}
\end{figure}

Figure~\ref{fig:Kepler_multiplicity} shows the results of a Monte Carlo experiment to determine the statistical robustness of this difference in multiplicity between the high and low-density \emph{Kepler} samples. We generate 1000 synthetic planet populations by randomly drawing from Poisson distributions with mean numbers of planets given by the observed numbers of single and multiple planet systems in the low and high-density sub-samples, and repeat the multiplicity analysis. The error bars in Figure~\ref{fig:Kepler_multiplicity} show the range from the 16th to 84th percentile of the Monte Carlo realisations. We repeat the multiplicity analysis using different probability thresholds that a given system is in a high-density ($P_{\rm high}$) or low-density ($P_{\rm low}$) environment.

Figure~\ref{fig:Kepler_multiplicity} shows that for planetary systems taken from a single observational survey (\emph{Kepler}) with a uniform selection criterion (i) there is a statistically significant difference between the single-to-multiple ratio between the high and low density samples, and (ii) that this difference increases as the probability that a given system is in a low or high-density environment increases. As the \emph{Kepler} sample contains few HJs \citep{wright12}, we conclude that the difference in multiplicity between low and high-density environments extends to the full planet population, and is not restricted to HJs alone. We also repeat the analysis using different cuts in stellar parameters to verify that the distributions of host star properties between the field and overdensities are statistically indistinguishable (see, e.g.\ the bottom rows of Table~\ref{tab:bias_check}).

In summary, no matter how the data are split, the same trend in planetary multiplicity between the low and high-density sub-samples is recovered, although the degree of statistical significance weakens with decreasing sample size.

\section{Discussion}
\label{sec:disc}

We now discuss what insight these observed differences in the field and overdensity planet populations may provide into the role that stellar clustering plays in the formation and evolution of planetary architectures. The planet multiplicity distribution around field stars provides a considerably better match to the multiplicity distribution of co-planar systems in the \citet{mulders18} models. Under the reasonable assumption that co-planar systems are the least likely to have been perturbed, the field systems represent an ideal sub-sample of exoplanetary systems to compare with simulations and models attempting to understand how planets form and evolve in effective isolation.

In contrast, we find a factor 1.6--2.0 excess of single-planet systems in overdensities compared to the field for the full \citet{w20nat} sample, and that this increases up to a factor 3.5 when only using planets detected by \emph{Kepler}.  This suggests that environments of high stellar phase space density play a prominent role in setting the planetary multiplicity, and may be a key contributing factor in the observed \emph{Kepler} dichotomy. Stellar clustering may therefore play an important role in creating the two different populations of planetary architectures that previous studies \citep[e.g.][]{lissauer11,hansen_murray13} have concluded are required to reproduce the observed exoplanet multiplicity distribution. Further  comparison between the overdensity and field planet populations offers a fruitful new avenue to distinguish the relative influence of the different mechanisms postulated to be responsible for producing the excess of single-planet systems \citep[e.g.][]{johansen12, hansen_murray13, morton_winn14, ballard16}. 

The observed trend that the orbital periods of single-planet systems in overdensities are smaller than the orbital periods of multiple-planet systems is consistent with previous studies showing that single-planet systems have shorter periods than systems hosting multiple planets \citep[e.g.][fig.~9]{weiss18b}. As this trend is far more pronounced in overdensities than in the field, whatever mechanism may be responsible for the period decrease in single-planet systems must be more effective when the host star resides in a higher density environment.

Having already concluded that entire planetary system architectures can be changed by the environment, the comparison between $\zeta_1$ and $\zeta_2$ for the field and overdensities probes the effect of the environment on mean motion resonances. As $\zeta_1$ between adjacent planet pairs is indistinguishable between the field and overdensities, first-order mean motion resonances must reform easily after disruption by environmental perturbations. Although the result is currently statistically marginal, the fact that the orbital period ratios of adjacent planets in the field may lie closer to second-order mean motion resonances (i.e.\ have lower $\zeta_2$ than overdensities) suggests that second-order resonances may reform less easily after environmental perturbations. Following this logic, second-order resonances could therefore either be imprints of the planet formation process, i.e.\ once they are destroyed they are not re-established, or they are never truly stable to perturbations and represent a transient state. In the latter case, the ratio between their formation and disruption timescales is higher in environmentally-perturbed systems than in environmentally unperturbed systems, causing their incidence to decrease.

Finally, we note a recent study by \citet{adibekyan21}, who compare the orbital period distributions between overdensities and the field identified by \citet{w20nat} for a small, but homogeneous sample of RV-detected planets with improved host stellar parameters, finding no significant difference. This result is consistent with our findings (as well as Extended Data Figure~8 of \citealt{w20nat}), which show that (1) RV detections are generally too small to identify statistically significant differences and (2) the difference between overdensities and the field is largest for transit detections. Our statistical tests using the California-Kepler Survey sample \citep{fulton_petigura18} corroborate this interpretation, demonstrating that the environmental dependence of multiplicity is not dominated by the Jupiter-mass planets detected in RV surveys.

In summary, the above analysis of planetary multiplicity and orbital period distributions as a function of host stellar phase space density shows that ambient stellar clustering in the large-scale environment plays an important role in shaping the architecture of planetary systems. Understanding the demographics of the planet population at large will require linking the physical mechanisms acting on this wide variety of different scales, from planet formation and evolution to stellar dynamics and galaxy evolution.

\vspace{3cm}

\acknowledgments
We thank the anonymous referee for insightful comments and suggestions which substantially improved the paper. We thank Paola Caselli, Trevor David, Laura Kreidberg, and Diego Turrini for helpful suggestions. M.C.\ and J.M.D.K.\ gratefully acknowledge funding from the Deutsche Forschungsgemeinschaft (DFG, German Research Foundation) through an Emmy Noether Research Group (grant number KR4801/1-1) and the DFG Sachbeihilfe (grant number KR4801/2-1). J.M.D.K.\ gratefully acknowledges funding from the European Research Council (ERC) under the European Union's Horizon 2020 research and innovation programme via the ERC Starting Grant MUSTANG (grant agreement number 714907). This research made use of data from the European Space Agency mission \textit{Gaia} (\href{http://www.cosmos.esa.int/gaia}{http://www.cosmos.esa.int/gaia}), processed by the \textit{Gaia} Data Processing and Analysis Consortium (DPAC, \href{http://www.cosmos.esa.int/web/gaia/dpac/consortium}{http://www.cosmos.esa.int/web/gaia/dpac/consortium}). Funding for the DPAC has been provided by national institutions, in particular the institutions participating in the \textit{Gaia} Multilateral Agreement. This research has made use of the NASA Exoplanet Archive, which is operated by the California Institute of Technology, under contract with the National Aeronautics and Space Administration under the Exoplanet Exploration Program.

\software{
\package{matplotlib} \citep{hunter07},
\package{numpy} \citep{vanderwalt11},
\package{pandas} \citep{reback20},
\package{scipy} \citep{jones01},
\package{seaborn} \citep{waskom20}
}

\bibliographystyle{aasjournal}
\bibliography{mybib}

\end{document}